\documentstyle[11pt,aaspp4,epsf]{article}
\include{psfig}

\newcommand{\Hi}{\mbox{H{\sc i}}}

\newcommand{\mic}{\mbox{$\mu$m}}

\newcommand{\sbs}{\mbox{SBS\,0335-052}}

\received{Sept. 3. 1998}
\accepted{Nov. 4, 1998}

\slugcomment{To appear in ApJ}

\lefthead{Thuan, Sauvage, \& Madden}
\righthead{ISOCAM observations of SBS 0335-052}

\begin{document}

\title{Dust in an extremely metal-poor galaxy: mid-infrared
observations of SBS 0335-052\footnote{Based on data obtained with ISO,
an ESA project with instruments funded by the ESA member states
(especially the PI countries: France, Germany, the Netherlands, and
the United Kingdom) with the participation of ISAS and NASA.}}

\author{Trinh X. Thuan}
\affil{Astronomy Department, University of Virginia, P.O. Box 3818,
University station, Charlottesville, VA 22903-0818; txt@virginia.edu}

\and

\author{Marc Sauvage and Suzanne Madden}
\affil{CEA/DSM/DAPNIA/Service d'Astrophysique, CE-Saclay, 91191 Gif
sur Yvette cedex, France; msauvage@cea.fr, smadden@cea.fr}

\begin{abstract}

The metal deficient (Z = Z$_{\odot}$/41) Blue Compact Dwarf Galaxy
(BCD) \sbs\ was observed with ISOCAM between 5 and 17\,\mic.  With a
L$_{12{\mu}m}$/L$_{B}$ ratio of 2.15, the galaxy is unexpectedly
bright in the mid-infrared for such a low-metallicity object.  The
mid-infrared spectrum shows no sign of the Unidentified Infrared
Bands, which we interpret as an effect of the destruction of their
carriers by the very high UV energy density in \sbs. The spectral
energy distribution (SED) is dominated by a very strong continuum
which makes the ionic lines of [S{\sc iv}] and [Ne{\sc iii}] very
weak.  From 5 to 17\,\mic\, the SED can be fitted with a grey-body
spectrum, modified by an extinction law similar to that observed
toward the Galactic Center, with an optical depth of
A$_{V}\,\sim$\,19-21 mag.  Such a large optical depth implies that a large
fraction (as much as $\sim$ 75\%) of
the current star-formation activity in \sbs\ is hidden by
dust with a mass between 3$\times$10$^{3}$\,M$_{\odot}$ and
5$\times$10$^{5}$\,M$_{\odot}$. 
Silicate grains are present as
silicate 
extinction bands at 9.7 and 18 \mic\ can account for the unusual 
shape of the MIR spectrum of \sbs.
 It is remarkable that such a nearly
primordial environment contains as much dust as galaxies which are 10
times more metal-rich.  If the hidden star formation in \sbs\ is
typical of young galaxies at high redshifts, then the cosmic star
formation rate derived from UV/optical fluxes would be underestimated.

\end{abstract}


\section{Introduction}

Galaxy formation is one of the most fundamental unsolved problems in
astrophysics.  Much effort has gone into the search for primeval
galaxies (hereafter PGs), i.e.  galaxies undergoing their first major
burst of star formation, at redshifts larger than $\sim$2.  Several
objects have been put forward as possible PGs, ranging from
high-redshift radio-galaxies to Ly$\alpha$ emitters
 (e.g. Steidel et al. 1996; Yee et al. 1996).
 However, most of these candidate PGs appear
to already contain a substantial amount of heavy elements (as
indicated for example by the presence of P Cygni profiles), implying
previous star formation and metal enrichment.

\sbs\ ($\alpha$(1950) = 03$^{h}$\,35$^{m}$\,15.2$^{s}$, 
$\delta$(1950) = $-05^{\circ}\,12'\,25.9\arcsec$)
 is a relatively nearby Blue Compact Dwarf Galaxy (hereafter BCD)
with an absolute magnitude M$_B$= -16.7 and 
 which appears to be undergoing its first
 burst of star formation.  With
a metallicity of only Z$_\odot$/41 (Izotov et al. 1997), it is the second
most metal-deficient galaxy known, after I Zw 18 (Z$_\odot$/50).  With
a redshift of 0.0136 and a Hubble constant of 75 km s$^{-1}$
Mpc$^{-1}$, it is at a distance of 54.3 Mpc (1$\arcsec$\,=\,263 pc).
Thuan, Izotov, \& Lipovetsky (1997) and \cite{papaderos98} have found
 that the stars
in \sbs\ are not older than $\sim$ 100 Myr, making it a truly young
galaxy.  \cite{thiso97} suggest that the large HI envelope within
which \sbs\ is embedded may be truly primordial.  Yet, despite its
youth and extremely low metallicity, HST images of \sbs\ clearly
show dust patches mixed in with the six super-star clusters  where most of
the star formation is occurring (Thuan et al. 1997).  The presence of
dust, in combination with the intense ultraviolet radiation
field from the many young stars in \sbs\, suggests that there
may be detectable Mid-Infrared (MIR) emission, as the dust will
reprocess the UV starlight and reemit it in the infrared.  We have
therefore obtained MIR observations of \sbs\ with the 
Infrared Space Observatory (ISO) to study the
properties of dust and star formation in a truly metal-deficient
environment, similar to those prevailing at the epoch of galaxy
formation.

\section{Observations and Data Reduction}
\label{sec:obs}

The observations were obtained with ISOCAM (Cesarsky et al., 1996a),
the MIR imaging instrument aboard ISO (Kessler et al., 1996).  They
consist of two different sets: a set of broad-band filter maps, and a
set obtained by imaging spectroscopy.

The broad-band maps were obtained with a spatial sampling of 3$''$ per
pixel and in a 3$\times$3 raster mode with a 3-pixel displacement in
each direction, giving a total field of view of $114''\times114''$.
The raster mode increases the sensitivity and improves the flat-field
correction since in the center of the map, each sky position is imaged
by 9 different camera pixels.  The maps were obtained with the LW9,
LW10, LW8, LW6 and LW2 filters, centered respectively at 14.9, 12.0,
11.3, 7.7 and 6.7\,\mic\ (Table~\ref{tbl:obslog}).  Note that the LW10
filter is identical to the IRAS 12\,\mic\ filter.  The integration
time was 10.08\,s per readout for all filters, except LW10 for which
an integration time of 5.04\,s was used.  Total on-source times were
860\,s for LW10 and 1540\,s for the other filters.


Imaging spectroscopy was performed using the Circular Variable Filter
(CVF) facility of ISOCAM. This mode produces images of the
full ISOCAM field of view ($192''\times192''$) in the 2-17\,\mic\ wavelength
region, with a sampling of 6$''$ per pixel and a spectral
resolution $\lambda$/$\Delta\lambda$ of $\simeq$ 40.  Due to the
faintness of the source we obtained observations only in the 8.78-17.34\,\mic\
wavelength range.  The individual integration time was
5.04\,s and the total on-source time was 4800\,s.

The various steps of ISOCAM data reduction are detailed in
\cite{starck98} and we will only describe here aspects that depart
from the methods discussed in that paper: dark current subtraction and
transient correction.

The dark current is known to show secular trends which depend on
the position of the satellite in its orbit and on the time elapsed
since launch.  These trends can be accurately
modelled (see Biviano et al. 1998),  allowing nominal dark subtraction.

Transient corrections using the inversion algorithm of Abergel et
al. (1996) were applied successfully to the broad-band observations.
For the CVF observations, the known oversimplification of the method
combined with the intrinsic faintness of the source results in errors
of similar magnitude for both corrected and uncorrected CVF data.  The
main effect of the correction is to add an artifact at the start of
the spectrum (around 17\,\mic) while leaving the rest of the scan
unchanged.  In particular, synthesis of the LW8 and LW9 flux densities
from corrected or uncorrected spectra gives flux densities 1.3 times
larger than the observed broad-band flux densities.  That the CVF
produces higher flux densities than the broad-band filters is to be
expected: the operational setup of ISOCAM is such that the scan starts
with a short exposure through the LW2 filter with a scale of 6$''$ per
pixel, giving a much higher illumination on the detector than the scan
itself. As the source is very faint, the detector would need a much
longer time than our adopted exposure time to stabilize down to its
true level. This is not the case for the broad-band maps as the
broader filters allow many more photons to reach the detector, thus
speeding the stabilization process.  As a result, we have adopted for
the remaining discussion the CVF spectrum uncorrected for transient
effects and scaled down by a factor of 1.3\,.

Figure 1 displays the spectral energy distribution of \sbs.  The
flat-fielding and photometry of the CVF scan were made following
\cite{agb98}.  Conversion from camera units to mJy was performed using
the calibration factors given in the {\em ISOCAM
Cookbook}\footnote{available at {\em
http://isowww.estec.esa.nl/manuals/iso\_cam/}\,.}.  HST images (Thuan
et al. 1997) reveal that most of the star formation in \sbs\ occurs
within a region of $\sim$2.5$''$ in size, so that the individual
super-star clusters in \sbs\ are not resolved by ISOCAM. Thus the flux
densities given in Table~\ref{tbl:obslog} and plotted in Figure 1 are
integrated over the whole star-forming region.


\section{Discussion}
\label{sec:disc}

\subsection{Global properties}

Since the ISOCAM LW10 filter is equivalent to the IRAS 12\,\mic\
filter, it is interesting to see how the integrated properties of
\sbs\ compare with those of other BCDs and irregular galaxies. We
have thus compiled IRAS and H$\alpha$ fluxes for the Thuan \& Martin
(1981, hereafter TM81) catalog of BCDs, as well as for the sample of BCDs
and irregular galaxies in Sauvage, Thuan, \& Vigroux (1990) to perform 
the comparison.

With a 12\,\mic\ to blue luminosity ratio L$_{12}$/L$_{B}$ of
2.15$\pm$0.06, \sbs\ is comparable to the most MIR-bright objects in
the TM81 catalog.  In that sample the mean L$_{12}$/L$_{B}$ ratio
($<$log\,L$_{12}$/L$_{B}$$>$\,=\,-0.12$\pm$0.27), while already twice
as high as that for spiral galaxies in the CfA catalog (Sauvage \&
Thuan 1994), is a factor of 3 smaller than the ratio for \sbs.  This
shows that even though the BCD is $\sim$ 10 times more metal-poor than
the galaxies in TM81, it nevertheless contains a significant amount of
dust heated by an intense UV radiation field.

If we compare the 12\,\mic\ luminosity to a tracer of star formation
such as the H$\alpha$ luminosity (Table~\ref{tbl:data}), \sbs\
appears to be normal.  Although total H$\alpha$ fluxes are scarce for
BCDs, the available data give
$<$log\,L$_{12}$/L$_{H\alpha}$$>$\,=\,0.96$\pm$0.70 for the above
samples, as compared to 0.65$\pm$0.03 for \sbs.  Thus, even though the
metallicity of the BCD is exceptionally low, and the galaxy unusually
MIR-bright, its star-forming properties appear to be normal when
compared to other BCDs.

\subsection{The mid-infrared spectrum}

When the MIR spectral energy distribution of \sbs\
(Figure 1) is compared to MIR spectra of other
star-forming galaxies (see e.g. Metcalfe et al. 1996; Vigroux 1997),
two facts stand out : (1) there is no emission from the so-called
Unidentified Infrared Bands (hereafter UIBs, usually attributed to
carbon-based dust, see e.g. Papoular et al. 1996; Puget \& L\'eger 1989).
In particular the very strong band at 11.3\,\mic\ is completely absent
from the spectrum; and (2) there are no evident fine structure ionic
lines, even though neon as well as sulfur lines are usually quite
bright in starburst galaxies.

In Figure 1, a hint of the [S{\sc iv}]$\lambda$10.5\mic\
and [Ne{\sc iii}]$\lambda$15.6\mic\ lines can be seen.  There is a
feature near the position of the [Ne{\sc ii}]$\lambda$12.8\mic\ line,
but it is centered at 12.8\,\mic, while the redshifted line falls at
12.98\,\mic.  The quality of our spectrum prevents us from actually
measuring line fluxes. Instead we compute upper limits as twice
the local rms noise times the instrumental profile width at 20\% of
peak intensity at the line
location.  We obtain upper limits of
5.6$\times$10$^{-14}$\,erg\,s$^{-1}$\,cm$^{-2}$ for the [S{\sc iv}]
line, 4.5$\times$10$^{-14}$\,erg\,s$^{-1}$\,cm$^{-2}$ for the [Ne{\sc
ii}] line, and 5.4 $\times$10$^{-14}$\,erg\,s$^{-1}$\,cm$^{-2}$ for
the [Ne{\sc iii}] line.

Using the photoionization models of \cite{stasinska90} with a
metallicity equal to 0.02 that of the sun, and with an integrated
H$\beta$ flux in \sbs\ of
12.1$\times10^{-14}$\,erg\,s$^{-1}$\,cm$^{-2}$ (the flux given by
Izotov et al. 1997 was multiplied by 4 to account for aperture and
extinction effects as suggested by Thuan \& Izotov 1997), we predict the
line intensities of [S{\sc iv}], [Ne{\sc iii}], and [Ne{\sc ii}] to be
5.0\,$\times$10$^{-14}$ erg\,s$^{-1}$\,cm$^{-2}$,
1.6\,$\times$10$^{-14}$\,erg\,s$^{-1}$\,cm$^{-2}$, and
7.7$\times$10$^{-17}$\,erg\,s$^{-1}$\,cm$^{-2}$ respectively.  The
predicted intensities are thus completely consistent with the upper
limits.  We conclude, therefore, that the weakness (or absence) of ionic
fine-structure lines in \sbs\ when compared to other star-forming
galaxies can be explained by a very strong continuum which decreases
the equivalent widths of the lines, making them difficult to detect.

None of the usual UIB features are seen, suggesting that their
carbonaceous carriers are absent or have been destroyed.  As the
star-forming region, 660\,pc in diameter, contains 5000 equivalent O7
stars (Izotov et al.  1997), the average energy density is
$\gtrsim$\,10\,eV\,cm$^{-3}$.  \cite{pugetleger89} suggest that at
this radiation density, the abundance of the band-emitting dust is
reduced by a factor of $\sim$10.  Therefore the absence of the bands
can be understood as a destruction effect.  This is probably enhanced
by the very low metallicity of the galaxy which allows harder UV
photons to travel further into the interstellar medium.

The origin of the continuum emission is more problematic.  We observe
what could be interpreted as either a broad emission feature at around
14\,\mic, or a broad absorption feature at $\lambda\gtrsim$\,16\,\mic.
We favor the second hypothesis as we know of no candidate which would
show such an emission feature at $\sim$ 14 \mic.  Very small
carbonaceous grains, usually thought to be responsible for the MIR continuum
(D\'esert, Boulanger, \& Puget 1990), produce featureless emission.
Silicates, the other well-known component of dust, can produce
emission features of various shapes, but these are centered around
10\,\mic\, and even in circumstellar disks never extend beyond
12\,\mic\ (see e.g.  Sitko et al.  1998).  On the other hand, silicate
extinction bands at 9.7 and 18\,\mic\ can provide an explanation for
the strange shape of the MIR spectrum of \sbs.

Testing this hypothesis is not straightforward as the expected shape
of the dust spectrum in this wavelength range is difficult to
constrain.  Apart from the stochastic heating of macromolecules which
produce the UIBs, MIR emission is thought to be produced by very
small grains in a thermal regime intermediate between equilibrium and
stochastic heating (see e.g.  D\'esert et al.  1990).  We have
therefore attempted to fit the spectral energy distribution (SED) of
\sbs\ with different screen models where the emission spectrum was
successively: (1) a black-body of variable temperature modified by an
emissivity law of the form f$_\nu$\,$\alpha$\,B$_\nu$(T)\,$\nu^{1-2}$;
(2) a power-law spectrum; and (3) the continuum MIR spectrum of the
Galactic HII region M17 with UIBs and ionic lines removed (Cesarsky et
al.  1996b).  The free parameters of the underlying spectra include in
all cases a scaling factor, in addition to the temperature in case (1)
and the power-law slope in case (2).  A further uncertainty comes from
the extinction curve to be used.  We have tried two different laws,
the standard MIR extinction law derived by \cite{draineESA89}, and
that observed in the direction of the Galactic Center by
\cite{lutzetal96}.  The second law differs from the first by its much
higher A$_{\lambda}$/A$_{V}$ in the range 2-8\,\mic.  This difference
is probably due to the neglect of ice coating on grains in the
\cite{draineESA89} model.  The free parameter associated with the
extinction curve is simply the absolute extinction at a reference
wavelength.  Our fitting procedure searches for best fits to the CVF
scan data only.  There is no attempt to fit the broad-band data
outside of the CVF range.  However, a successful model will be one
that also accurately predicts them.


Figure 2 shows examples of the three types of fits.  Both
a power law and the M17 spectrum fail to reproduce the observed SED.
Power laws modified by extinction overproduce emission shortward of
8\,\mic\ (Figure 2a) while the M17 spectrum overestimates
emission longward of 14\,\mic\ (Figure 2b).  On the
contrary, a black-body spectrum modified by an emissivity law
proportional to $\nu^{1.5}$ and extinguished by a screen of dust,
gives an excellent fit to the SED of \sbs\ (Figure 2c).
The goodness of the fit depends very little on the exponent of the
emissivity law, so its value cannot be used to constrain the
nature of the emitting dust.  Instead, the fit is sensitive to the
shape of the extinction law.  It is impossible to obtain a good fit
for the LW2 and LW6 bands with the standard extinction curve of
\cite{draineESA89}.  The predicted spectrum always overproduce the
emission in these bands by factors of $\sim$2.  On the contrary,
excellent fits of these bands can be obtained with the \cite{lutzetal96}
 extinction curve.  The temperature obtained for the black-body
curve is in the range 240-260 K, but probably carries little physical
meaning since the MIR-emitting grains are not likely to be in thermal
equilibrium.

The information we derive concerning the dust extinction is more
physical.  Given the uncertainties in our data, we obtain an
equivalent A$_{V}$ of 19-21 mag.  Note that since the screen model
maximizes A$_{V}$ for a given column density of dust, the
column density derived from our fit is likely a lower limit to
the true column density if emitting and absorbing dust grains are
mixed.

Such a high value of extinction is in sharp contrast to the values
(A$_{V}$ $\leq$ 0.6 mag) derived by \cite{izotov97} from optical
spectrophotometric observations.  The large difference between the
optical and infrared extinctions implies that the heating sources
responsible for the infrared emission detected by ISO are probably too
deeply embedded in dust to be detected in the visible and that, as in
the case of the Antennae galaxies NGC\,4038/39 (Mirabel et al.  1998),
most of the infrared emission is powered by invisible star clusters.
Indeed, correcting the H$\alpha$ flux for 20\,mag of visible
extinction would lead to unrealistically high star formation rates
which would be incompatible with the non-detection of \sbs\ at
1.4\,GHz by the NVSS (Condon et al.  1998), or in the far-infrared by
IRAS.

The LW10 luminosity of 2.5\,$\times$10$^{8}$\,L$_{\odot}$ is
equivalent to the bolometric luminosity of 2500\,O7 stars, i.e.  half
the equivalent number of stars in the visible super-star clusters.
Given that in the BCD sample of TM81, $<$log\,L$_{12}$/L$_{\rm
FIR}$$>$\,=\,-0.85$\pm0.09$, it is probable that in fact some 18000
equivalent O7 stars are required to power the total infrared emission.
This implies that even in very low-metallicity environments such as
those characterizing PGs, a significant fraction, i.e.  $\sim$ 3/4, of
the total star formation activity of a galaxy can be effectively
hidden from UV/optical observations.

From the extinction derived in the MIR, we can estimate the dust mass.
As MIR extinction is essentially due to silicates (Draine \& Lee 1984),
this result
will likely be a lower limit as some carbon-based dust (such as the
grains responsible for the continuum) can be present and still
contribute little extinction in the MIR. Using the dust model of
\cite{dl84}, we obtain a dust surface density of 1.5
M$_{\odot}$\,pc$^{-2}$.  To derive a dust mass, we need to estimate
lower and upper bounds to the dust spatial extent.  For a lower bound,
we assume that the dust is only associated with hidden super-star
clusters (SSCs).  O'Connell, Gallagher, \& Hunter (1994) found that SSCs
typically have a very compact core with a half-light radius of
$\sim$\,3\,pc, embedded in considerably more extended halos with
diameters $\sim$\,30\,pc.  If we adopt the latter value, and consider
that at least 3 SSCs are hidden (they have the power of 2500\,
equivalent O7 stars needed to power the LW10 luminosity), we derive a
dust mass of $\sim$\,3.2\,$\times$\,10$^3$\,M$_{\odot}$.  This is very
probably an underestimate of the true dust mass as there are likely
many more hidden SSCs.  Furthermore dust is probably not just
associated with the SSCs, but rather mixed throughout the star-forming
region, especially if it has a supernova origin as argued later on.

A likely upper limit to the spatial extent of the dust is
the size of the region where reddening is observed by HST (Thuan et
al.  1997), i.e.  660\,pc.  This gives a dust mass of
5$\times$10$^{5}$\,M$_{\odot}$.  With the \Hi\ mass given in
Table~\ref{tbl:data}, the gas-to-dust mass ratio of \sbs\ is then
between $\sim\,2\,\times\,10^{3}$ and $\sim\,3\,\times\,10^{5}$, i.e.
much higher than in the Galaxy, which is not too surprising for such a
low metallicity object (Lisenfeld \& Ferrara 1998).  More
extraordinary is the fact that the dust mass in \sbs\ can be of the
same order of magnitude as that in BCDs which are on average $\sim$ 10
times more metal-rich.  For comparison
$<$log\,M$_{dust}$$>$\,=\,4.4$\pm$0.6\,M$_{\odot}$ in the TM81 BCD
sample.

As discussed by \cite{txtsbs97}, because stars in \sbs\ are not
older than $\sim$100 Myr, there is not enough time for the silicate
dust to be made in the envelopes of red giant stars.  Rather silicate
grains probably condensed out of the numerous supernova ejecta present
in the BCD (see e.g.  \cite{lucyetal91} on SN\,1987A; Dwek et al.
1992; Wooden et al. 1993).  We can check for the plausibility of this
hypothesis by adopting, for example, the supernova silicate 
dust mass input rate of 0.5\,M$_{\odot}$\,pc$^{-2}$\,Gyr$^{-1}$ 
obtained by \cite{dwek98}
for the Galactic Center at 0.1\,Gyr ,the age of
\sbs. Assuming the diameter of the dust-forming region to be 660\,pc,
 we derive a silicate dust mass
of $\sim$2$\times$10$^{4}$\,M$_{\odot}$, in the range of the above
estimates.

\section{Conclusions and Implications}

Although \sbs\ is one of the most metal-deficient galaxies known
($Z_{\odot}$/41), it is unexpectedly bright in the MIR range, implying
a silicate dust mass in the range 10$^3$--10$^5$ M$_\odot$.  The MIR
spectrum shows no sign of UIB carriers which are probably destroyed.
Despite a difference of a factor of 40 in
metallicity, the Galactic Center MIR extinction produces the best fit
to the spectrum. This is quite different from the situation in the UV
where the extinction law shows a strong dependence on metallicity
(Fitzpatrick 1989).  This difference is probably due to the fact that
the abundance and spectral properties of the carbon-based dust
responsible for the UV extinction and UIBs are more
metallicity-dependent than those of the silicate grains responsible
for the extinction in the MIR. A possible explanation is that the
silicate grains are more resistant to photo-destruction than
carbon-based dust.

The derived extinction is quite high, A$_V$\,$\sim$\,20 mag.  Given
that the total MIR luminosity already requires the bolometric
luminosity of 50\% more young stars than are seen in the galaxy,
the total star formation rate (SFR) as derived
from the optical or UV luminosities must underestimate the true star
formation rate by at least 50\%, and more likely by a factor of 4 as
argued above.  Thus if \sbs\ is a representative example of PGs, the
cosmic star formation rate will be systematically underestimated if based
only on UV and optical fluxes.  This is in fact
the result obtained by \cite{flores98} in their ISO survey of distant
galaxies.  They found that the cosmic SFR derived from FIR
luminosities is $\sim$ 2 to 3 times higher than the SFR estimated
previously from UV/optical fluxes (Madau et al. 1996).

\acknowledgements

The ISOCAM data was analyzed using the software package ``CIA'', a
joint development by the ESA Astrophysics Division and the ISOCAM
consortium.  The ISOCAM consortium is led by the ISOCAM PI, C.
Cesarsky, Direction des Sciences de la Mati\`ere, C.E.A, France.
T.X.T. has been partially supported by NASA grant JPL961535.  We acknowledge
useful conversations with Yuri Izotov.

\newpage



\begin{figure}
\plotone{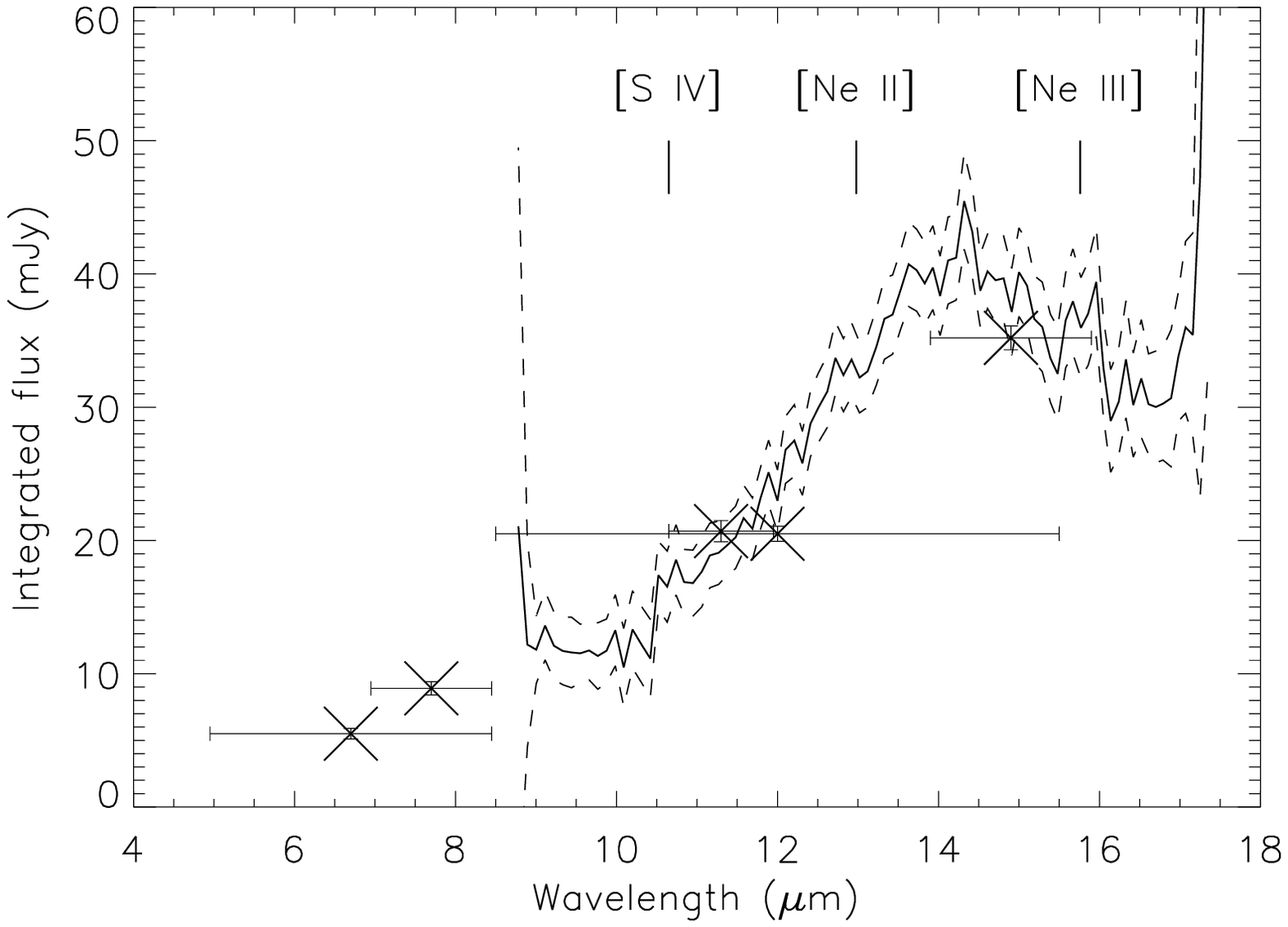}

\caption{Spectral energy distribution of SBS 0335-052 in the
mid-infrared.  The crosses represent broad-band observations at 6.7,
7.7, 11.3, 12.0 and 14.9 \mic.  The horizontal bars show the width of
the broad-band filters.  The 1$\sigma$ errors bars are smaller than
the crosses.  Broad-band flux densities are derived assuming an
intrinsic spectral shape $f_{\nu}\,\propto\,\nu^{-1}$ and integrating
over the point spread function.  The thick line is the CVF spectrum
without transient correction and scaled down by a factor 1.3 (see text
for details).  The dashed lines show $\pm\,1\sigma$ errors.  The
spectrum has been corrected for effects of the varying point spread
function with wavelength.  Expected positions of the [S{\sc
iv}]$\lambda$10.5\mic, [Ne{\sc ii}]$\lambda$12.8\mic, and [Ne{\sc
iii}]$\lambda$15.6\mic\ emission lines are indicated.}

\label{fig:sed}
\end{figure}

\begin{figure}
\epsscale{0.5}
\plotone{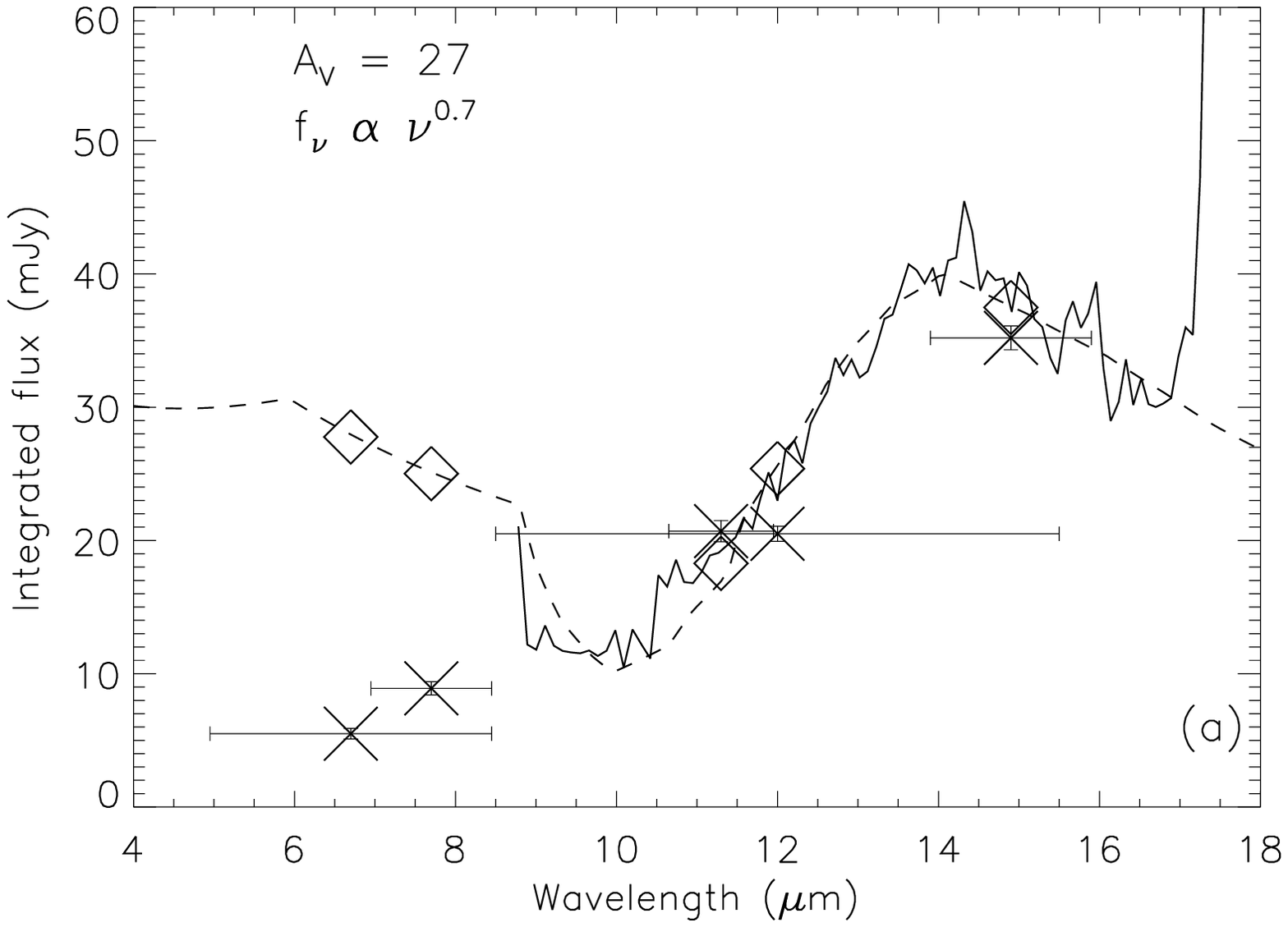}
\\
\plotone{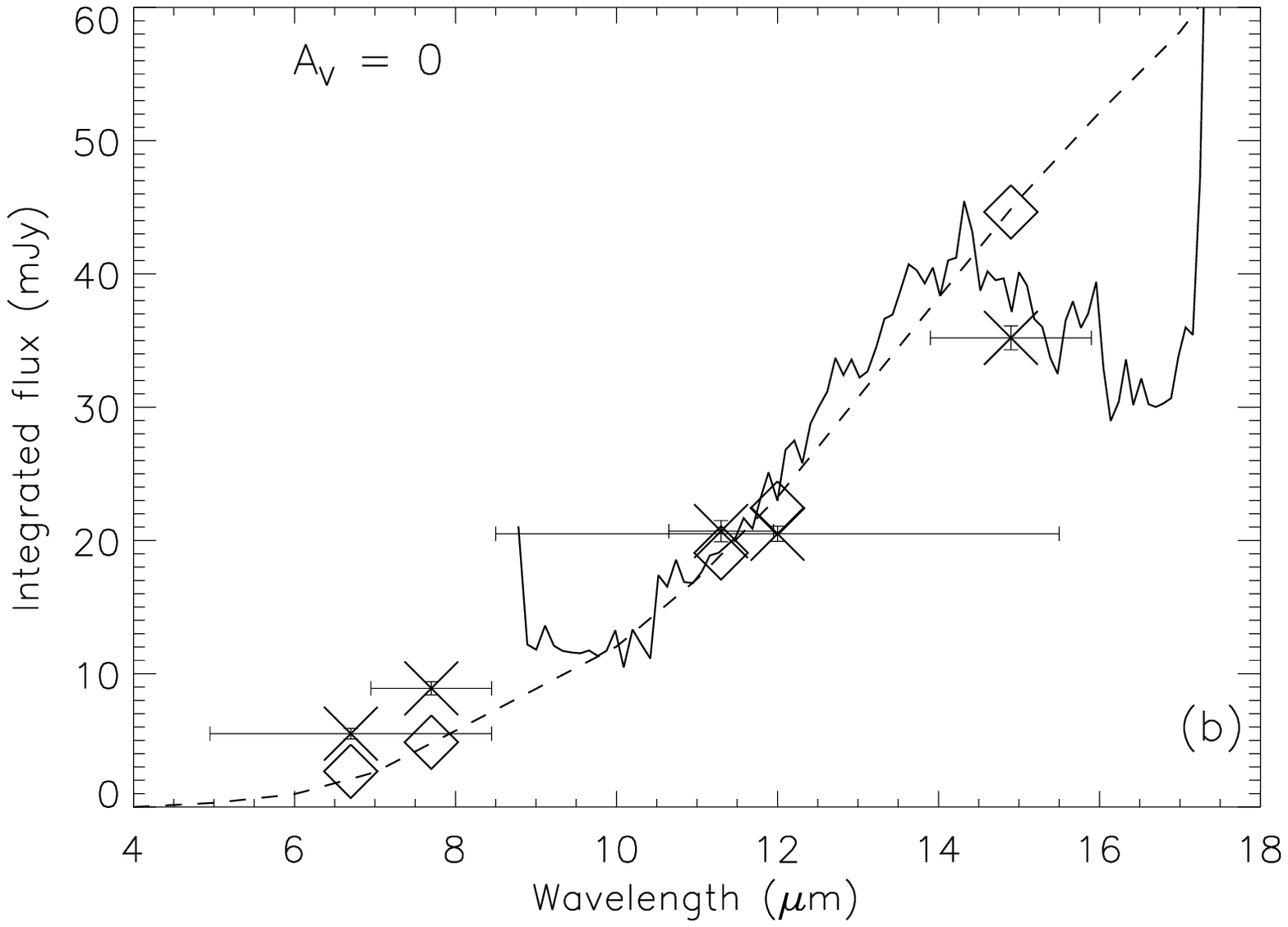}
\\
\plotone{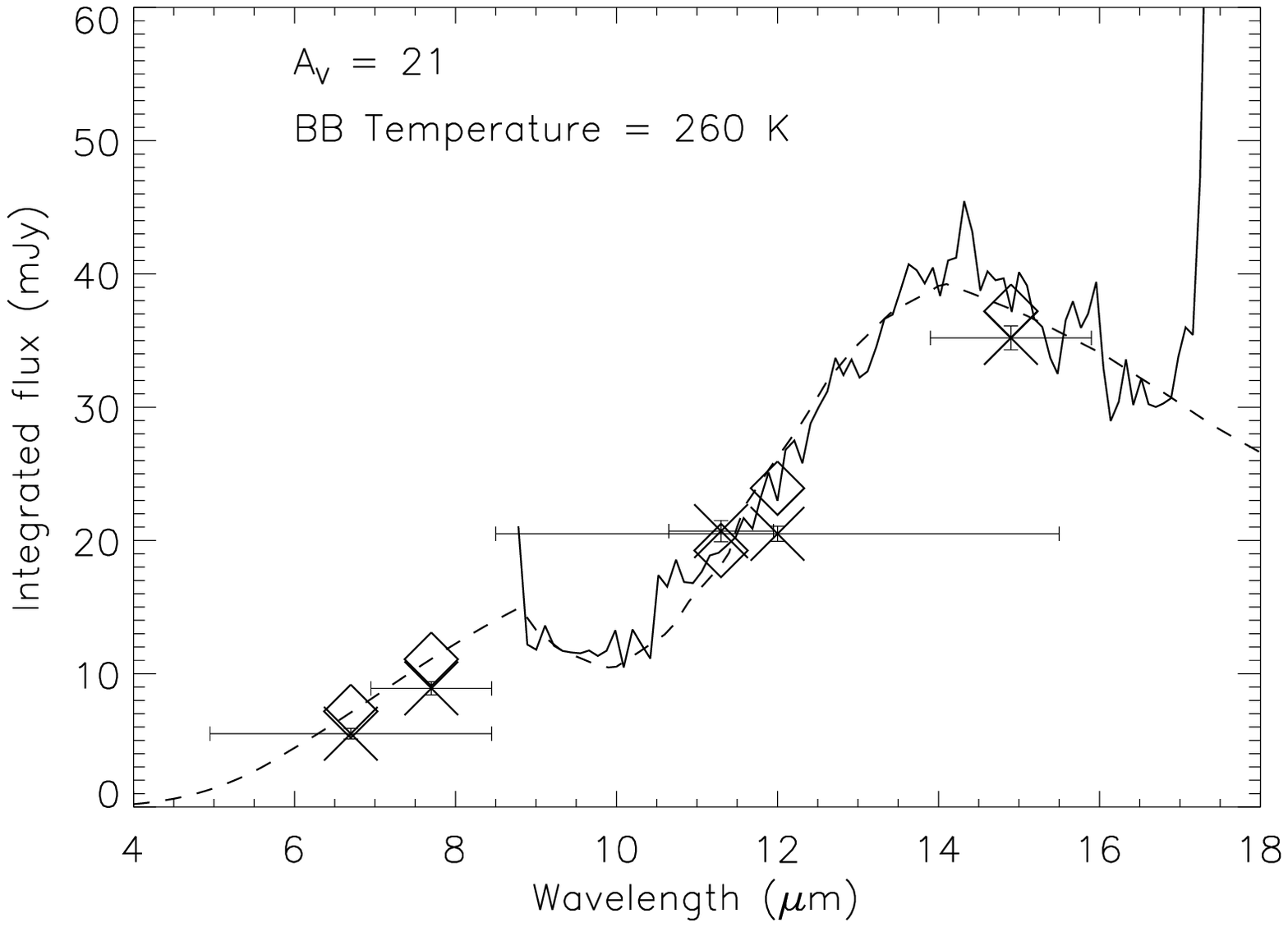}

\caption{Best-fit models (dashed lines) for the spectral energy
distribution of \sbs.  Three different types of emission spectra are
considered: (a) a power-law of the form $\nu^{0.7}$ with an extinction
A$_{V}$\,=\,27 mag; (b) the continuum observed in M\,17 (Cesarsky et
al.  1996b) with A$_{V}$\,=\,0; (c) a black-body of temperature 260\,K,
with an exponent of 1.5 for the emissivity law, and A$_{V}$\,=\,21\,
mag.  For all three fits the extinction curve of \cite{lutzetal96}
gives the best results.  To compare our fit to the broad-band data
(crosses), we have integrated the fitted spectrum (dashed line) over
the band-pass of the ISOCAM filters.  The resulting flux densities are shown
by open diamonds.  Only model (c) is able to reproduce satisfactorily
all broad-band observations.}

\label{fig:fit}
\end{figure}


\begin{deluxetable}{lrrrcrr}
\tablecaption{MIR flux densities of SBS 0335-052\label{tbl:obslog}}

\tablehead{
\colhead{filter} & \colhead{$\lambda_{0}$\tablenotemark{a}} &
\colhead{$\Delta\nu$\tablenotemark{a}} & 
\colhead{$\Delta\lambda$\tablenotemark{a}} & \colhead{Flux density} &
\colhead{1$\sigma$} & \colhead{Luminosity\tablenotemark{b}} \\
\colhead{} & \colhead{\mic} & \colhead{10$^{12}$\,Hz} & \colhead{\mic} &
\colhead{mJy} & \colhead{mJy} & \colhead{L$_{\odot}$}}

\startdata
LW9  & 14.9 &  2.29 & 2.0  & 35.2 & 0.9 & 7.4\,10$^{7}$ \nl
LW10 & 12   & 13.20 & 7.0  & 20.5 & 0.6 & 2.5\,10$^{8}$ \nl
LW8  & 11.3 &  2.71 & 1.3  & 20.7 & 0.8 & 5.2\,10$^{7}$\nl
LW6  & 7.7  &  7.06 & 1.5  & 8.9  & 0.5 & 5.8\,10$^{7}$\nl
LW2  & 6.7  & 16.18 & 3.5  & 5.5  & 0.4 & 8.2\,10$^{7}$\nl
\enddata
\tablenotetext{a}{Assuming a spectral shape
$f_{\nu}\propto\nu^{-1}$, as in the ISOCAM cookbook.}
\tablenotetext{b}{Assuming a distance of 54.3\,Mpc.}
\end{deluxetable}

\newpage

\begin{deluxetable}{ll}
\tablecaption{Relevant data for \sbs\label{tbl:data}}

\tablehead{\colhead{Parameter} & \colhead{Value}}

\startdata
L$_{H\alpha}$\tablenotemark{a}      &   5.6\,10$^{7}$\,L$_{\odot}$\nl
L$_{H\beta}$\tablenotemark{a}       &   2.2\,10$^{7}$\,L$_{\odot}$\nl
L$_{B}$\tablenotemark{b}            &   1.2\,10$^{8}$\,L$_{\odot}$ \nl
L$_{B}$\tablenotemark{c}            &   7.2\,10$^{8}$\,L$_{B}^{\odot}$ \nl
M$_{\rm H{\sc i}}$\tablenotemark{d} &   9.5\,10$^{8}$\,M$_{\odot}$ \nl
\enddata
\tablenotetext{a}{Flux in the 1\arcsec\ slit of Izotov et al. 1997,
corrected for
extinction using A$_{V}$\,=\,0.6 mag, and multiplied by 2 to account
for aperture effects as suggested by Thuan \& Izotov 1997.}
\tablenotetext{b}{Computed from the absolute blue magnitude in Thuan
et al. 1997.}
\tablenotetext{c}{Here L$_{B}$ is expressed in units of the solar blue
luminosity, where L$_{\odot}$/L$_{B}^{\odot}\,\sim$\,6.25\,.}
\tablenotetext{d}{Thuan et al. 1998.}
\end{deluxetable}


\newpage
\begin{thebibliography}{}


\bibitem[Abergel et al. (1996)]{abergelOph96}
Abergel, A., et al.  1996,  \aap, 315, L329

\bibitem[Aussel et al. (1998)]{agb98}
Aussel, H., Gerin, M., Boulanger, F., D\'esert, F. X., Casoli, F.,
Cutri, R. M., \& Signore, M. 1998,  \aap, 334, L73

\bibitem[Biviano et al. (1998)]{dkmod98}
Biviano, A., Sauvage, M., Roman, P., Boulade, O., Gallais, P., \&
Okumura, K. 1998, ``The ISOCAM dark current calibration report'', ESA
Technical report

\bibitem[Cesarsky et al. (1996a)]{ccisocam96}
Cesarsky, C., et al. 1996a, \aap, 315, L32

\bibitem[Cesarsky et al. (1996b)]{cesrskm17}
Cesarsky, D., Lequeux, J., Abergel, A., Perault, M., Palazzi, E., Madden, S.,
\& Tran, D.  1996b,  \aap, 315, L309

\bibitem[Condon et al. (1998)]{condonetal98}
Condon, J. J., Cotton, W. D., Greisen, E. W., Yin, Q. F., Perley, R. A.,
Taylor, G. B., \& Broderick, J. J.  1998,  \aj, 115, 1693

\bibitem[D\'esert et al. (1990)]{dbp90}
Desert, F. X., Boulanger, F., \& Puget, J. L.  1990,  \aap, 237, 215

\bibitem[Draine (1989)]{draineESA89}
Draine, B. T.  1989,  in Infrared spectroscopy in astronomy, ed. B. H.
Kaldeich,
(Noordwijk: ESA), 93

\bibitem[Draine \& Lee (1984)]{dl84}
Draine, B. T., \& Lee, H. M.  1984,  \apj, 285, 89

\bibitem[Dwek et al. (1992)]{dweketal92}
Dwek, E., Moseley, S. H., Glaccum, W., Graham, J. R., Loewenstein, R.
F., Silverberg, R. F., \& Smith, R. K. 1992, \apj, 389, L21

\bibitem[Dwek (1998)]{dwek98}
Dwek, E. 1998, \apj, 501, 643

\bibitem[Fitzpatrick (1989)]{fitzp89}
Fitzpatrick, E. L.  1989,  in Interstellar dust,
ed. L. J. Allamandola, \& A. G. G. M. Tielens (Dordrecht: Kluwer), 37

\bibitem[Flores et al. (1998)]{flores98}
Flores, H., et al.  1998,  \apj, submitted

\bibitem[Izotov et al. (1997)]{izotov97}
Izotov, Y., Lipovetsky, V. A., Chaffee, F. H., Foltz, C. B., Guseva,
N. G., \& Kniazev, A. Y.  1997,  \apj, 476, 698

\bibitem[Kessler et al. (1996)]{kessleriso96}
Kessler, M. F., et al. 1996, \aap, 315, L27

\bibitem[Lisenfeld \& Ferrara (1998)]{lisenfeld98}
Lisenfeld, U., \& Ferrara, A.  1998,  \apj, 496, 145

\bibitem[Lucy et al. (1991)]{lucyetal91}
Lucy, L. B., Danziger, I. J., Gouiffes, C., \& Bouchet, P.  1991,
in Supernovae, ed. S. E. Woosley (New York: Springer), 82

\bibitem[Lutz et al. (1996)]{lutzetal96}
Lutz, D., et al.  1996,  \aap, 315, L269

\bibitem[Madau et al. (1996)]{madauetal96}
Madau, P., Ferguson, H., Dickinson, M., Giavalisco, M., Steidel, C.,
\& Fruchter, A.  1996,  \mnras, 283, 1388

\bibitem[Metcalfe et al. (1996)]{metcalfe96}
Metcalfe, L., et al.  1996,  \aap, 315, L105

\bibitem[Mirabel et al. (1998)]{mvc98}
Mirabel, F., Vigroux, L., Charmandaris, V., Sauvage, M., Gallais, P.,
Cesarsky, C., Madden, S., \& Duc, P.A. 1998, \aap, 333, L1

\bibitem[O'Connell et al. (1994)]{oconnell94}
O'Connell, R. W., Gallagher, J. S., \& Hunter, D. A.  1994,  \apj, 433, 65

\bibitem[Papaderos et al. (1998)]{papaderos98}
Papaderos, P., Izotov, Y. I., Fricke, K. J., Thuan, T. X., \& Guseva,
N. G. 1998, \aap, 338, 43

\bibitem[Papoular et al. (1996)]{pap96}
Papoular, R., Conard, J., Guillois, O., Nenner, I., Reynaud, C.,
\& Rouzaud, J. N.  1996,  \aap, 315, 222

\bibitem[Puget \& L\'eger (1989)]{pugetleger89}
Puget, J. L., \& Leger, A.  1989,  \araa, 27, 161

\bibitem[Sauvage \& Thuan (1994)]{st94}
Sauvage, M., \& Thuan, T. X.  1994,  \apj, 429, 153

\bibitem[Sauvage et al. (1990)]{stv90}
Sauvage, M., Thuan, T. X., \& Vigroux, L.  1990,  \aap, 237, 296

\bibitem[Sitko et al. (1998)]{sitko98}
Sitko, M. L., Grady, C. A., Lynch, D. K., Russel, R. W., \& Hanner, M.
S. 1998, \apj, submitted

\bibitem[Stark et al. (1998)]{starck98}
Starck, J. L., et al.  1998,  \aap, in press

\bibitem[Stasinska (1990)]{stasinska90}
Stasinska, G.  1990,  \aaps, 83, 501

\bibitem[Steidel et al. (1996)]{Steidel96b}
Steidel, C. C., Giavalisco, M., Pettini, M., Dickinson, M.,
\& Adelberger, K. L.  1996,  \apjl, 462, L17

\bibitem[Thuan \& Martin (1981)]{tm81}
Thuan, T. X., \& Martin, G. E.  1981,  \apj, 247, 823

\bibitem[Thuan \& Izotov (1997)]{thiso97}
Thuan, T. X., \& Izotov, Y. I.  1997,  \apj, 489, 623

\bibitem[Thuan et al. (1997)]{txtsbs97}
Thuan, T. X., Izotov, Y. I., \& Lipovetsky, V. A.  1997,  \apj, 477, 661

\bibitem[Thuan et al. (1998)]{txt98}
Thuan, T. X, Lipovetsky, V. A., Martin, J. M., \& Pustilnik, S. A. 1998,
\aaps,
in press

\bibitem[Vigroux (1997)]{vigroux97}
Vigroux, L. 1997, in Extragalactic astronomy in the infrared, ed. G.
Mamon, T. X. Thuan, \& J. T. T. Van, (Paris: Editions Fronti\`eres), 63

\bibitem[Wooden et al. (1993)]{woodenetal93}
Wooden, D. H., Rank, D. M., Bregman, J. D., Witteborn, F. C.,
Tielens, A. G. G. M., Cohen, M., Pinto, P. A., \& Axelrod, T. S. 1993,
\apjs, 88, 477

\bibitem[Yee et al. (1996)]{yee96}
Yee, H. K. C., Ellingson, E., Bechtold, J., Carlberg, R. G.,
\& Cuillandre, J. C.  1996,  \aj, 111, 1783
\end{thebibliography}
\end{document}